\documentclass[prl,epsfig,floats,twocolumn,superscriptaddress,amssymb,amsmath,floatfix,showpacs]{revtex4-1}  
\usepackage{graphicx}
\usepackage{bm}  
\usepackage{amssymb}
\usepackage{color}
\usepackage{amscd}

\begin{document}

\title{The Boson Peak and Disorder in Hard Sphere Colloidal Systems}

\author{Rojman Zargar}
\affiliation{Institute of Physics, University of Amsterdam, Science Park 904, 1098 XH Amsterdam, the Netherlands}
\author{John Russo}
\affiliation{Institute of Industrial Science, University of Tokyo, 4-6-1 Komaba, Meguro-ku, Tokyo 153-8505, Japan}
\author{Peter Schall}
\affiliation{Institute of Physics, University of Amsterdam, Science Park 904, 1098 XH Amsterdam, the Netherlands}
\author{Hajime Tanaka}
\affiliation{Institute of Industrial Science, University of Tokyo, 4-6-1 Komaba, Meguro-ku, Tokyo 153-8505, Japan}
\author{Daniel Bonn}
\affiliation{Institute of Physics, University of Amsterdam, Science Park 904, 1098 XH Amsterdam, the Netherlands}


\begin{abstract}
The Boson peak is believed to be the key to the fundamental understanding of the anomalous thermodynamic properties of glasses, notably the anomalous peak in the heat capacity at low temperatures; it is believed to be due to an excess of low frequency vibrational modes and a manifestation of the structural disorder in these systems.
We study the thermodynamics and vibrational dynamics of colloidal glasses and (defected) crystals. The experimental determination of the vibrational density of states allows us to directly observe the Boson peak as a strong enhancement of low frequency modes. Using a novel method [Zargar \textit{et al.}, Phys. Rev. Lett. \textbf{110}, 258301 (2013)] to determine the free energy, we also determine the entropy and the specific heat experimentally. It follows that the emergence of the Boson peak and high values of the specific heat are directly related and are specific to the glass: for a very defected crystal with a disorder that is only slightly smaller than for the glass, both the low-frequency density of states and the specific heat are significantly smaller than in the glass.
\end{abstract}
\maketitle
The vibrational density of states (DOS) and the normal modes of solids provide a direct route to study its thermodynamics and mechanical properties \cite{Kittel}. In perfect crystalline solids, because of their long range order, vibrational states are well understood as plane-wave phonon modes \cite{Aschcroft,AntinaPhysicaA,Gratale,ChenPRE2013}. However, for more disordered systems the nature of vibrations remains elusive \cite{WyartEPL2005}. Structurally disordered systems such as liquids, glasses and amorphous materials exhibit a number of peculiar properties that are anomalous compared to those of the crystals \cite{WyartEPL2005,AntinaPRL1,AntinaSoft,ChenPRL2010,Kaya,Chumakov}. These properties include anomalous acoustic behavior, a peak in the temperature dependence of the specific heat $C_p/T^3$, and a Boson peak observed in inelastic scattering of light or neutrons \cite{WyartEPL2005,AntinaPRL1,AntinaSoft,ChenPRL2010,Kaya,Chumakov}. These suggest the existence of an excess vibrational density of states over and above the predictions of the Debye model: at the maximum in $C_p/T^3$, the vibrational density of states, $D(\omega)$,
scaled with the DOS of a perfect crystal,
goes through a maximum which is called the 'Boson peak' \cite{Phillips,Frick,Angell,Greaves,Xu,Shintani}. There is a consensus that the Boson peak is a manifestation of structural disorder, but its physical origin has remained a serious puzzle in condensed matter physics \cite{Phillips,Frick,Angell,Greaves,Xu,Shintani,Anderson}.
\begin{figure}
\includegraphics[width=1.00\columnwidth]{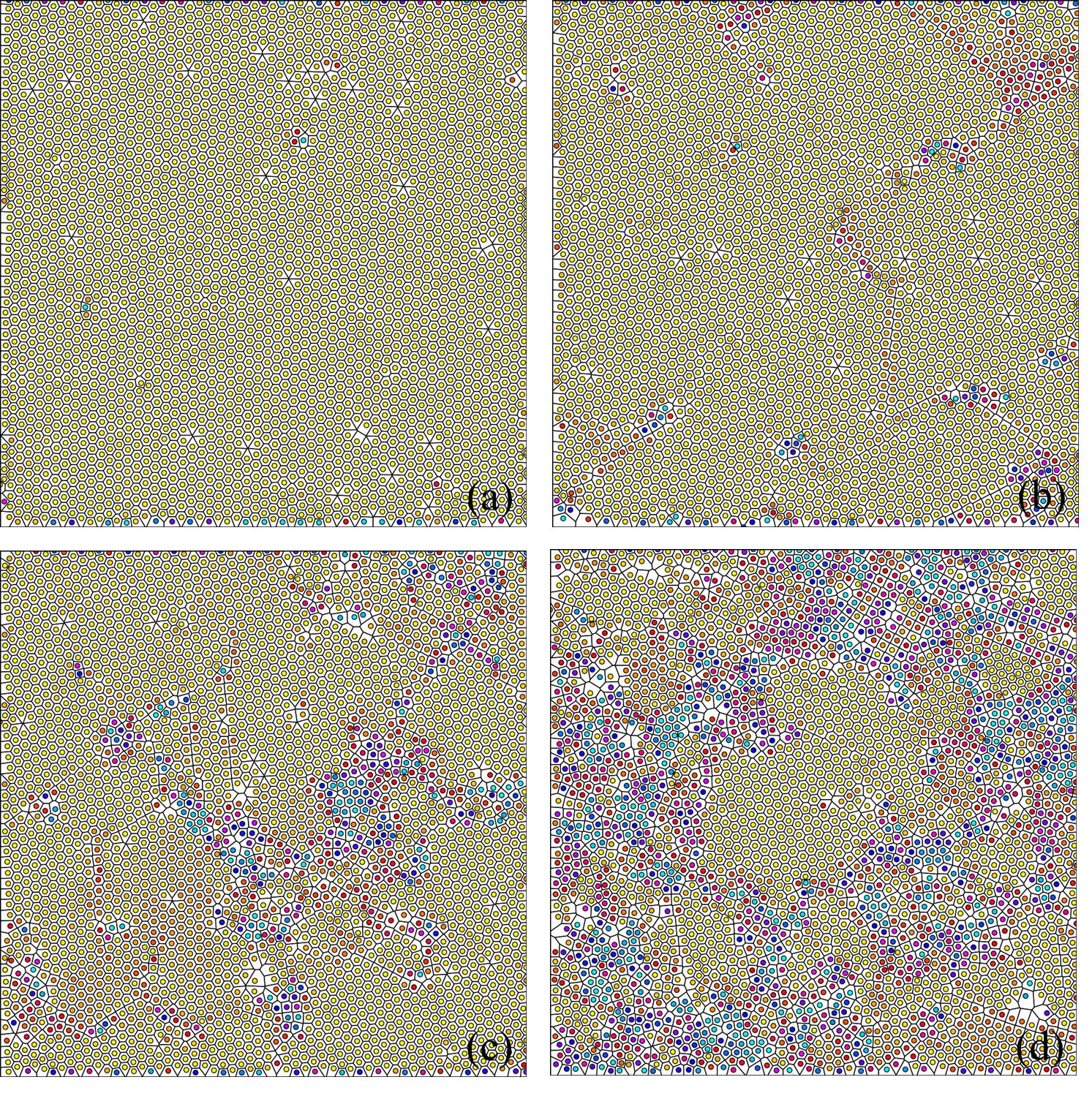}
\caption{Configuration of the particles in a 2D plane. The average position of the particles are shown for (a) a nearly perfect crystal, (b),(c) imperfect crystals, and (d) for a crystal involves a large amounts of disorder, a 'very defected crystal'. The amounts of disorder increases from (a) to (d). Particles are colored according to the hexatic order. A Voronoi tessellation is shown for each configuration.}
\label{Fig1}
\end{figure}
\begin{figure*}
\includegraphics[width=1.4\columnwidth]{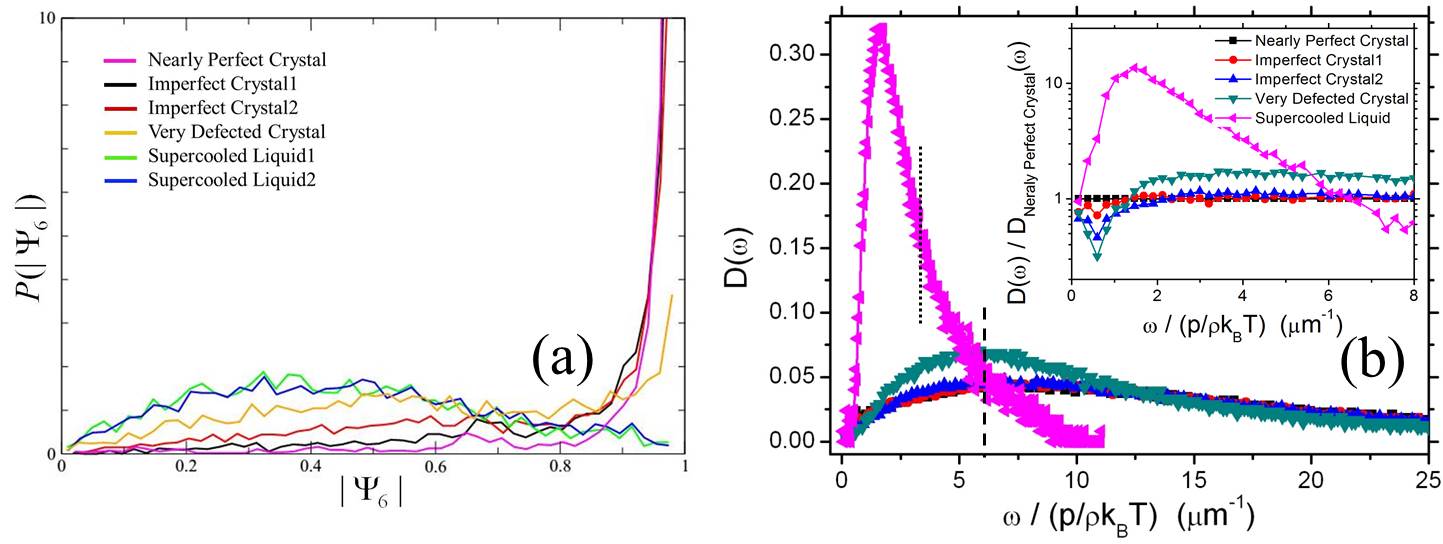}
\caption{(a) The probability density distribution of the bond orientational parameters for a nearly perfect crystal, crystals with defect, and for liquids. (b) The density of states for a nearly perfect crystal, crystals with three different amounts of disorder, and for a completely disordered system all at a same volume fraction $\phi=0.56$.
Dashed lines show the upper limit of the frequency until which the data is not affected by noise for crystals (at higher frequency) and for liquid. All the densities of states are scaled with the DOS of the nearly perfect crystal (Inset).}
\label{Fig2}
\end{figure*}
\par In this Letter, we study the effects of structural disorder on the vibrational modes and the thermodynamics of colloidal hard spheres. This system, as we will show below, allows us to determine both the DOS and the thermodynamic properties of the glassy and crystalline states, and to provide a direct comparison between the two states of matter.
We apply the covariance matrix analysis \cite{AntinaPRL1, Kurchan,Silke} to determine the density of states and the normal modes of vibrations from the particle displacements for a nearly perfect crystal, crystals with different amounts of defects, and for completely disordered systems (glasses).
We find that there is a strong enhancement of low frequency modes in the DOS for glasses which is signaling the Boson peak; however, no significant excess of modes is observed for very defected crystals, even though a quantification of the disorder shows that this is very similar between the glass and the disordered crystals.
We also experimentally determine the entropy, which is a measure of the specific heat at constant temperature, for several hard sphere systems with different amounts of disorder. We show that while the specific heat increases gradually with increasing amount of disorder for crystals, it shows a discontinuous jump between a very defected crystal and a glass, as does the intensity of the Boson peak. These observations confirm independently that the Boson peak is an intrinsic property of glassy systems and is not related to structural defects, since the very defected crystal is only slightly more ordered than the glass, i.e. the fraction of particles with local six-fold symmetry in the very defected crystal is only slightly larger than that of the glassy system.
\par Fast confocal microscopy allows us to determine the structure and dynamics of 1.5 $\mu m$ fluorescently labeled colloidal PMMA particles that are subject to thermal agitation (see the Supplementary Material for more details). For the DOS, we acquire sequences of 2D images on a 3D system; this allows us to follow the dynamics of the individual particles. The entire crystal is polycrystalline, allowing to perform the measurements on perfect and defected crystals on the same sample but at different regions that are characterized by a different defect density. For determining the free energy, we perform full 3D scans of the particle positions, that allow us to determine the free volumes of a given static configuration.
Figure \ref{Fig1} shows a 2D configuration of the particles for a nearly perfect crystal and crystals with three different amounts of disorder; crystals are more and more defected going from (a) to (d). The particles are colored according to their hexatic order; to quantify the order in our systems, we calculate the bond orientational order parameter,$\Psi_6=\frac{1}{N_{nn}}\sum_k^{N_{nn}}exp(6i\theta_{jk})$ in which $N_{nn}$ is the number of nearest neighbors and $\theta_{jk}$ indicates the direction of particle $j$ with respect to its nearest neighbors $k$ \cite{John}. Figure \ref{Fig2}(a) demonstrates the probability density distribution of $\Psi_6$. We find that for the nearly perfect crystal and the two more imperfect crystals, corresponding respectively to Fig. \ref{Fig1}(a),(b) and (c), the distribution shows a high peak at $\Psi_6=1$: all the particles are six-fold coordinated. For the glassy sample, the histogram is broad with $\Psi_6<1$. The very defected crystal however, represents both features: a small peak at $\Psi_6=1$ and a rather broad distribution for $\Psi_6<1$ (Figs. \ref{Fig1}(d) and \ref{Fig2}(a)).
\par Following the motion of around $2600$ particles in real time, we obtain all particle positions $x=x(t)$, $y=y(t)$ as functions of time using standard particle tracking software \cite{Crocker}. Denoting $u_i(t)$ the components of the particle displacements from the average positions along the confocal plane $u_i(t)=\{(x_i(t)-<x_i>), (y_i(t)-<y_i>)\}$, we obtain the displacement correlation matrix (of dimension twice the number of particles) as:
\begin{equation}
D_{lm}=<u_{\mu i}u_{\nu j}>, ~~\mu, \nu =x, y
\end{equation}
where $l,m=1,2,...,2N$ matrix index on the left runs both over the particle indices and the Cartesian components of displacements. The averaging $<...>$ has been done over the period of measurement, which is about 220 seconds.
\par Diagonalizing $D_{lm}$ we obtain the eigenvalues, $\lambda_m$, and the corresponding $2N$ normal modes of the system. Results are presented in terms of the mode frequencies which are related to the eigenvalues as:
\begin{equation}
\omega_m=\sqrt{1/\lambda_m}.
\end{equation}
\par The resulting density of states, $D(\omega)$, is shown in Fig. \ref{Fig2}(b); the DOS is plotted versus the frequency for a nearly perfect crystal, crystals with three different amounts of disorder and for a supercooled liquid all at a same volume fraction $\phi=0.56$. Since 'hard' modes are expected to have eigenvalues proportional to the pressure, we scale out this effect by plotting the density of states in terms of the scaled frequency $\omega/p$ \cite{AntinaPRL1,AntinaSoft,WyartPRE2005}; to do so, we use the Hall \cite{Hall} equation of state for the crystal and  the Liu \cite{Liu} equation of state for the supercooled liquid and glassy phases. The density of states is normalized such that $\int _0^\infty D(\omega)=1$. It has been established \cite{Silke} that for high frequencies the experimental noise becomes important due to the lack of accuracy in determining individual particle positions. Dashed lines show the limits below which the frequencies should remain unaffected by noise (Fig. \ref{Fig2}(b)).
\begin{figure}
\includegraphics[width=0.95\columnwidth]{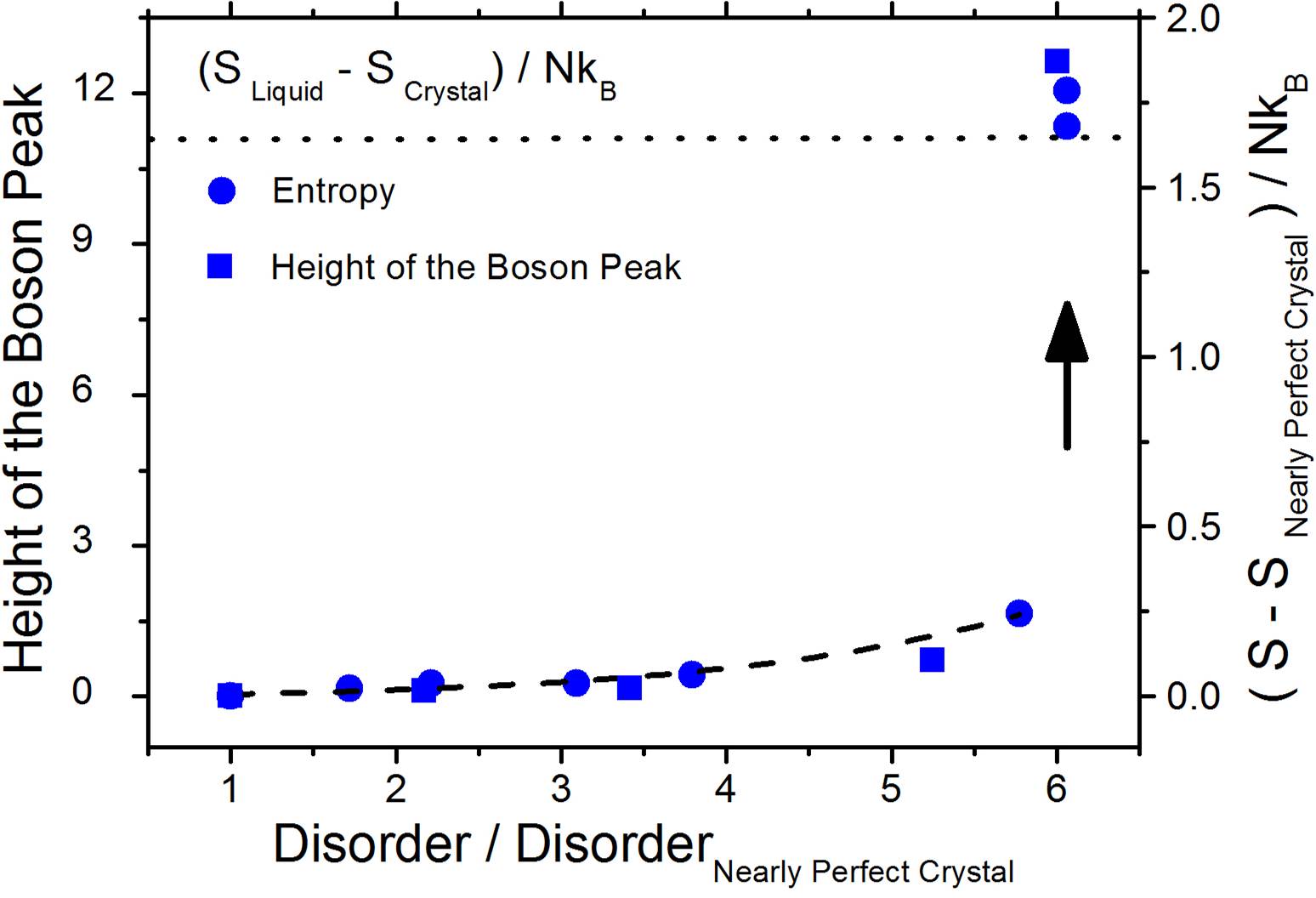}
\caption{Height of the Boson peak (Y-axis at left) is plotted versus the disorder for a nearly perfect crystal, crystals with three different amounts of disorder and for a supercooled liquid all at a same volume fraction $\phi=0.56$ (squares). The entropy per particle, which is indeed a measure of the specific heat for a system of hard spheres at constant temperature, in units of $k_B$ (Y-axis at right) is plotted versus the disorder for a nearly perfect crystal, crystals with four different amounts of disorder and for two supercooled liquids all at a same volume fraction $\phi=0.56$ (circles). Disorders are scaled with respect to that in the nearly perfect crystal for both.
Dotted line shows the difference between the entropy of a crystal and a liquid calculated in \cite{Freeenergy}. Dashed curve is plotted as a guide for eye.}
\label{Fig3}
\end{figure}
Figure \ref{Fig2}(b) shows that for a given low frequency, the observed density of states for defected crystals is larger than the DOS for a perfect crystal; the difference increases with increasing disorder. For the supercooled liquid however, the difference is much larger than any of the crystal samples (Fig. \ref{Fig2}(b)). We also find that the DOS shows a shift towards lower frequencies with increasing disorder. The most striking observation is however the large difference between all crystalline samples and the supercooled liquid, which shows a large peak at low frequencies.
\par To investigate the effects of disorder more consistently, we normalize all the densities of states with respect to that of the nearly perfect crystal (Fig. \ref{Fig2}(b) inset).
The very existence of the peak in the density of states for supercooled liquid with respect to that of a perfect crystal implies a strong excess of low frequency modes: this is the Boson peak.
We find that the excess of low frequency modes or equivalently, the height of the peak is very small for defected crystals compared to completely disordered systems (Fig. \ref{Fig2}(b) inset), implying that the Boson peak is an intrinsic property of glasses and does not arise due to structural disorder in crystals.
\par To check whether the DOS for a very defected crystal represents more crystalline features or is more liquidlike, we measure the density of states for several crystals which involve a large amounts of defects, i.e. with distribution of the $\Psi_6$ lies between crystal and liquid, (e.g. Figs. \ref{Fig1}(d) and \ref{Fig2}(a)); we find that for all of them, the observed DOS stays much closer to that of the perfect crystal and far below the DOS for the liquids (see Fig. 3S, Supplementary Materials), implying that there is a  discontinuous jump between the vibrational density of states for a very defected crystal and that of a supercooled liquid or a glass. This happens in spite of the observation that the fraction of particles with local six-fold symmetry in the very defected crystals is only slightly larger than in a glassy system. 
\begin{figure*}
\includegraphics[width=1\columnwidth]{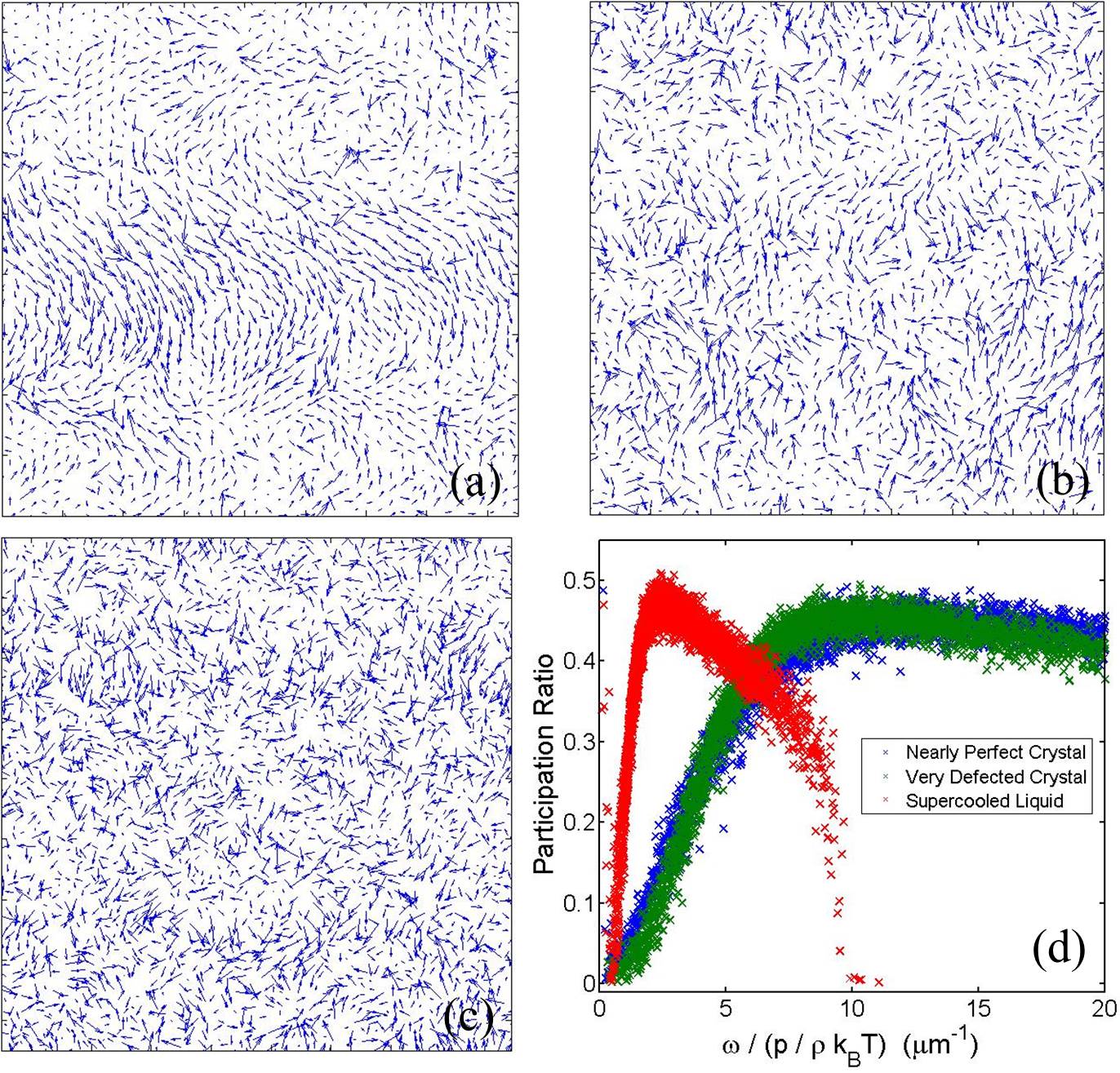}
\caption{Spatial distributions of the normal modes at $\omega/p=0.37$ for (a) a nearly perfect crystal, (b) a very defected crystal and (c) a supercooled liquid. (d) Participation ratio versus the frequency for a nearly perfect crystal, a very defected crystal and for a supercooled liquid all at a same volume fraction $\phi=0.56$.}
\label{Fig4}
\end{figure*}
\par We proceed by investigating the disorder dependence of the Boson peak in more detail (Fig. \ref{Fig3}). For each system we quantify the disorder exploiting two quantities: 1) the bond order parameter and 2) the number of nearest neighbors; a particle is considered as a defect if the former is less than 0.95 or the latter differs from six, the expected value for a 2D crystalline lattice. Disorder is then defined as the fraction of defected particles relative to the total number of particles, i.e. it is zero for a perfect crystal and one for a completely disordered system.
We find that, although the amounts of disorder in a supercooled liquid is only slightly larger than that in a very defected crystals however, the height of the Boson peak for the former is significantly larger, while it is negligible for the latter (Fig. \ref{Fig3}).
\par A unique feature of hard sphere colloids is that due to the absence of interactions, the free energy can be obtained directly from the configuration of the particles \cite{Dullens}. To study the effects of disorder on the thermodynamics, we then measure the entropy for several 3D subsystems involving different amounts of defects \cite{Freeenergy}; each subsystem that we measure contains around 2500 particles.  We calculate the free energy $F$ , from which the entropy follows directly as $F/T$. We determine the free volume for each individual particle; once the free volume is known, the free energy can be obtained directly from the cell model \cite{Freeenergy}. For a system of hard spheres at constant temperature $T$, any changes in the entropy indicates a change in the specific heat according to: $S(T)=\int{C_p(T) \frac{dT}{T}}$.
Results are shown in Fig. \ref{Fig3} where the entropy is plotted versus the disorder.
Interestingly, we find that, similarly to the height of the Boson peak, the specific heat for crystals increases gradually with increasing disorder and shows a significant jump between a very defected crystal and a supercooled liquid (Fig. \ref{Fig3}).
We observe again that while the difference between the amount of disorder for the very defected crystal and the supercooled liquid is very small, the difference between their measured entropy is very large, implying that the thermodynamic first order phase transition between crystal and liquid is not much affected by the disorder. This observation confirms also that the Boson peak is a property of glassy systems and is not related to structural disorder in crystals.
\par We now explore the effects of disorder on the normal modes of the different systems. Figure \ref{Fig4}(a),(b), and (c) shows the spatial distribution of the normal modes at a low frequency for respectively a nearly perfect crystal, a very defected crystal, corresponding to Fig. \ref{Fig1}(a),(d), and a supercooled liquid.
We find that the lowest frequency modes in the crystals exhibit mostly plane wave-like features that extend over very large length scales. This becomes less and less evident  with increasing the amount of disorder (Fig. \ref{Fig4}(a),(b)).
For the completely disordered system, the very lowest frequency modes show spatially correlated motions only over a few particle diameters.
\par To take a closer look at the nature of the modes, we compute the participation ratio which is a measure of the degrees of spatial localization of the modes. Modes are normalized for each of frequencies, so that $\sum_{i=1}^N \nu_i(\omega).\nu_i(\omega)=1$. The participation ratio is then defined as:
\begin{equation}
p(\omega)=(\sum_i \nu_i(\omega)^2)^2/ (N \sum_i \nu_i(\omega)^4),
\end{equation}
where $\nu_i(\omega)$ is the projection of the normal mode of frequency $\omega$ onto a colloidal particle $i$. Figure \ref{Fig4}(d) shows the participation ratio versus the frequency for a nearly perfect crystal, a very defected crystal and for a supercooled liquid. For the crystals, while most of the modes have participation ratios near $0.5$, the value expected for a plane wave, some of the low-frequency modes have a significantly smaller participation ratio, implying that for some low frequency modes, only a small fraction of the particles participates in the mode; these modes are known as quasi-localized modes \cite{ChenPRL2010,Tan}. The participation ratio for a nearly perfect crystal is typically larger than for a very defected crystal, indicating lower spatial localization of the modes for a nearly perfect crystal. For disordered systems however, the number of modes with low participation ratio is much larger than those in crystals, indicating a larger number of quasi-localized modes for disordered systems.
\par  In summary, we present the first experimental evidence that the Boson peak is an intrinsic property of glasses and does not arise due to structural disorder in crystals.
We show that the vibrational and thermodynamical properties, the DOS and the entropy, for a defected crystal are significantly different from those for a supercooled liquid or a glass, implying the different nature of the glass compared to a defected crystal. The large difference between the DOS and the thermodynamic properties of the defected crystalline and glassy systems then indicates that it is not possible to form a glass just by increasing the amount of disorder in a crystal, in spite of the fact that it may not be evident telling these systems apart experimentally by quantifying the disorder.
\par  We thank M. Schindler, A. C. Maggs for helpful discussions. We would like to thank the Stichting voor Fundamenteel Onderzoek der Materie (FOM) and Shell for financial support. This study was partly supported by Grants-in-Aid for Scientific Research (S) and Specially Promoted Research from JSPS.

\end{document}


\title{Supplementary Material for ''The Boson Peak and Disorder in Hard Sphere Colloidal Systems''}

\author{Rojman Zargar}
\affiliation{Institute of Physics, University of Amsterdam, Science Park 904, 1098 XH Amsterdam, the Netherlands}
\author{John Russo}
\affiliation{Institute of Industrial Science, University of Tokyo, 4-6-1 Komaba, Meguro-ku, Tokyo 153-8505, Japan}
\author{Peter Schall}
\affiliation{Institute of Physics, University of Amsterdam, Science Park 904, 1098 XH Amsterdam, the Netherlands}
\author{Hajime Tanaka}
\affiliation{Institute of Industrial Science, University of Tokyo, 4-6-1 Komaba, Meguro-ku, Tokyo 153-8505, Japan}
\author{Daniel Bonn}
\affiliation{Institute of Physics, University of Amsterdam, Science Park 904, 1098 XH Amsterdam, the Netherlands}

\maketitle
We use sterically stabilized fluorescent poly-methylmethacrylate particles suspended in a refractive index and density matched mixture of cis-decalin and cycloheptyl bromide which are the best model system for hard spheres \cite{Royall}.
Our particles have a diameter $\sigma=1.7 \mu m$ and polydispersity $\simeq 4\%$. The disordered system is obtained in a second system with $\sigma=1.5 \mu m$ and polydispersity $\simeq 7\%$ to prevent crystallization. In both systems an organic salt, tetrabutylammonium bromide, is used to screen any possible residual charges. The disordered systems at volume fractions between $0.54$ and $0.62$ and crystal at $0.56$ are prepared by diluting sediments that are centrifuged to random close packing ($\phi_{rcp}\simeq 0.64$) \cite{Poon}. In order to verify the hard sphere nature of the colloids the crystallization density, a very sensitive measure for deviations from the hard sphere behavior, is measured: the results agree to that of the true hard spheres within a fraction of a percent.
We use a fast confocal microscope (Zeiss LSM Live) to acquire 2D images of a 3D system in a field of view $105\mu m \times 105\mu m$. The 2D slices are taken at a distance of $25-30\mu m$ away from the coverslip, deep enough to avoid the effects of the boundary. The time interval between consecutive images is $0.04$ sec, which is approximately $0.024$ and $0.016$ of the Brownian time, $\tau _B=\eta_s d^3/k_BT$, of the particles in respectively the disordered system and the crystals. $\eta_s$ is the viscosity of the solvent.
\par
\begin{figure}
\includegraphics[width=1.05\columnwidth]{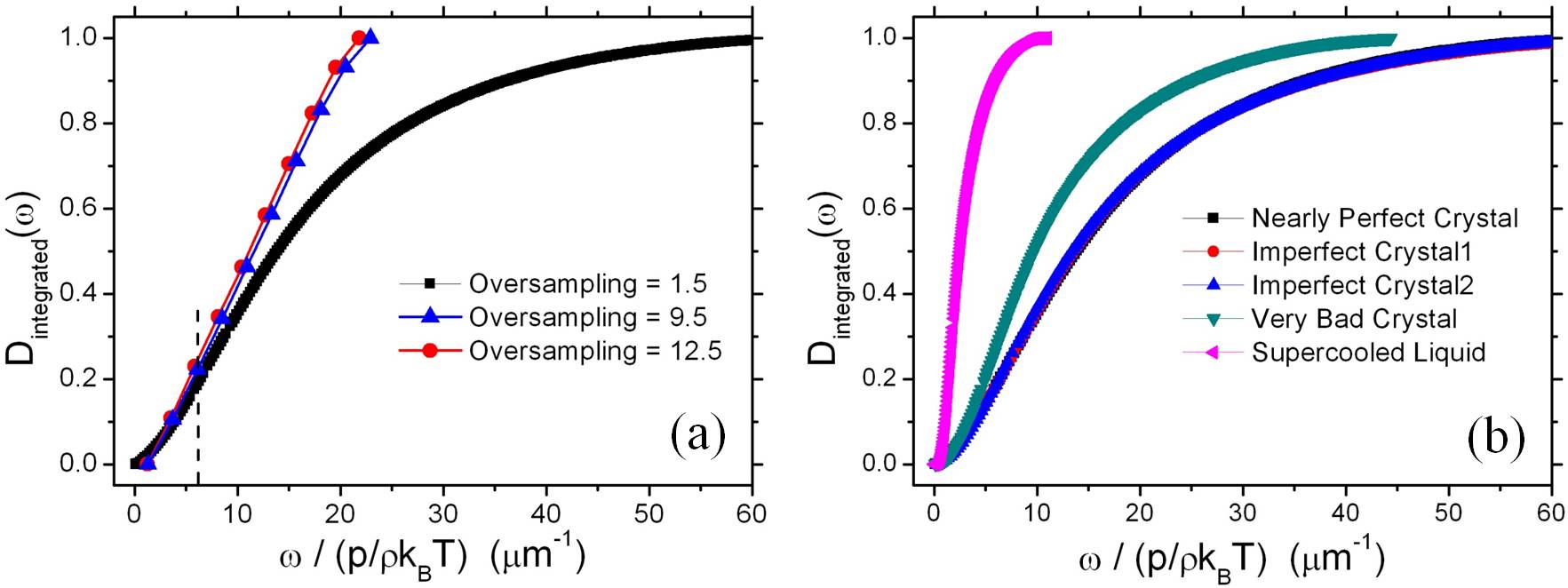}
\caption{(a) Cumulative density of states for nearly perfect crystals with different values for oversampling. (b) Cumulative density of states for a nearly perfect crystal, crystals with three different amount of disorder and for a supercooled liquid all at a same volume fraction $\phi=0.56$ and oversamplings around $\gamma\simeq2$. The frequency is scaled with the pressure which is obtained from the Hall and Liu equations of states for respectively crystal and liquid.}
\label{Fig1_Supp}
\end{figure}

  Measuring the density of states for a nearly perfect crystal, we identify a peak and a shoulder.
However, ideally, there are two peaks in the vibrational spectrum for a perfect crystal that are vestiges of van Hove singularities. Several factors may in our experiment contribute to the rounding of the van Hove singularities in our colloidal crystal, e.g. size polydispersity of particles, uncertainties in finding particle positions and the limitation due to the finite number of frames which all can introduce noise into the covariance matrix and thus into its eigenvalues and eigenvectors; on the other hand, at low frequencies the density of states is hardly affected by this. We investigate the effects of the finite number of statistics on the vibrational density of modes for a nearly perfect crystal. Figure \ref{Fig1_Supp}(a) shows the cumulative DOS for nearly perfect crystals with different values for oversampling. The oversampling is defined as ratio between the number of frames, $N_f$, and the number of degrees of freedom:
$\gamma=N_f/2N$, where $N$ is the number of particles.
We find that the low frequency part of the DOS spectrum remains unchanged with varying the oversampling by almost one order of magnitude (Fig. \ref{Fig1_Supp}(a)), indicating
our measured density of states for crystals and disordered system with oversamplings $\gamma\simeq2$ should not be affected by noise due to lack of statistics at low frequency.
\par Figure \ref{Fig1_Supp}(b) shows the cumulative DOS for a nearly perfect crystal, crystals with three different amounts of defects and for a supercooled liquid all at a same volume fraction $\phi=0.56$ and oversamplings $\gamma\simeq2$. We find that effects of structural disorder on the density of states is considerably larger than the effects of oversampling (Fig. \ref{Fig1_Supp}(a),(b)). The cumulative DOS shows a shift to lower frequencies with increasing the amount of disorder.
The strong enhancement of low frequency modes for the supercooled liquid is clear in figure \ref{Fig1_Supp}(b).
\par
\begin{figure}
\includegraphics[width=0.8\columnwidth]{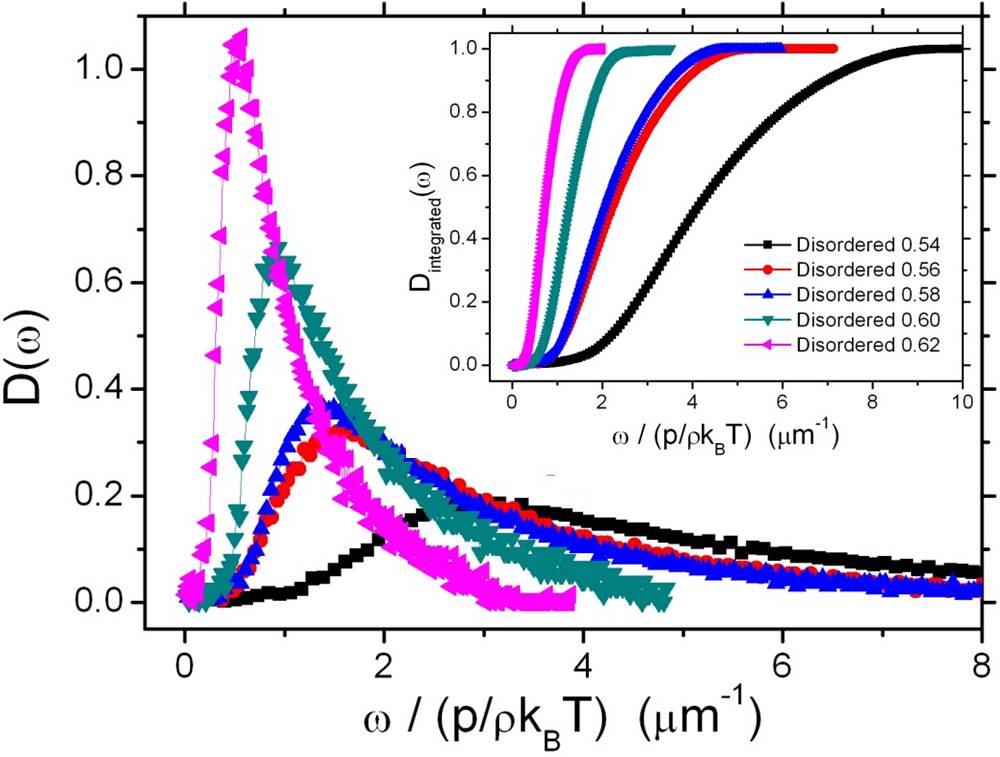}
\caption{Density of states for disordered system at different volume fractions $0.54$, $0.56$, $0.58$, $0.60$ and $0.62$. Frequencies are scaled with the dimensionless pressure in Liu equation of states. Dashed lines show the upper limit of the frequency until which the data is not affected by noise. Inset shows the cumulative DOS for different volume fractions.}
\label{Fig2_Supp}
\end{figure}

 We measure the density of states for completely disordered (liquid, glassy) systems at different volume fractions $0.54$, $0.56$, $0.58$, $0.60$ and $0.62$ (Fig. \ref{Fig2_Supp}).
For all volume fractions the DOS rises starting from a lowest frequency up to a maximum resulting a peak at a certain frequency and then starts decaying. Since the compression effect is taken out by scaling the frequency with the pressure, therefore by increasing the volume fraction, $D(\omega)$ shifts to lower $\omega/P$. The cumulative density of states for disordered system at different volume fractions is also shown in Fig. \ref{Fig2_Supp} inset.
\par
\begin{figure}
\includegraphics[width=1.0\columnwidth]{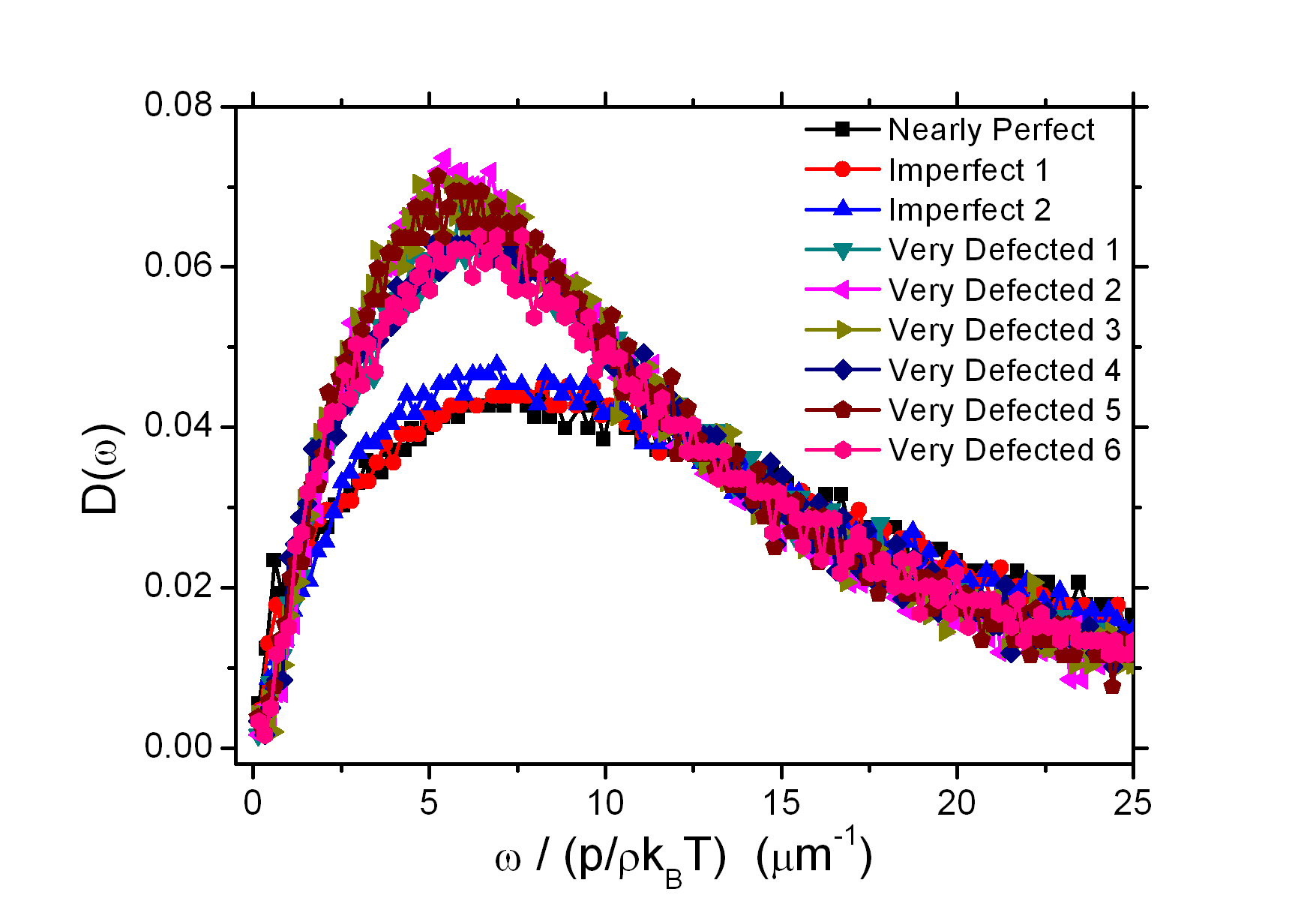}
\caption{Density of states for a nearly perfect crystal, two imperfect crystals and several very defected crystals.}
\label{Fig3_Supp}
\end{figure}

 To check whether the density of states for very defected crystals has more crystalline features or is more liquidlike, we measure the DOS for several crystals that involve large amounts of disorder, i.e. with the $\Psi_6$ distribution between crystals and liquids; we find that for all very defected crystals the observed density of states is much closer to that of the perfect crystal and far below the DOS for the liquids (Fig. \ref{Fig3_Supp}).
The fraction of particles with local six-fold symmetry in the very defected crystals is only slightly larger than in a completely disordered system; however, there is a significant difference between the DOS of the two, indicating the different nature for these systems.
